# Spontaneous atomic ordering and magnetism in epitaxially stabilized double perovskites

Akira Ohtomo[1,2,3*], Suvankar Chakraverty[4], Hisanori Mashiko[1], Takayoshi Oshima[1] and Masashi Kawasaki[5,4]

[1] *Department of Applied Chemistry, Tokyo Institute of Technology, Tokyo 152-8552, Japan.*

[2] *Tokodai Institute of Element Strategy (TIES) and Materials Research Center for Element Strategy (MCES), Tokyo Institute of Technology, Yokohama 226-8503, Japan.*

[3] *ALCA, Japan Science and Technology Agency (JST), Tokyo 102-0076, Japan.*

[4] *Correlated Electron Research Group (CERG) and Cross-Correlated Materials Research Group (CMRG), RIKEN Advanced Science Institute, Wako 351-0198, Japan.*

[5] *Quantum-Phase Electronics Center and Department of Applied Physics, The University of Tokyo, Tokyo 113-8656, Japan.*

*electronic mail: aohtomo@apc.titech.ac.jp

## Abstract

We have studied the atomic ordering of $B$-site transition metals and magnetic properties in the pulsed-laser deposited films of $La_2CrFeO_6$ (LCFO) and $La_2VMnO_6$ (LVMO), whose bulk materials are known to be single perovskites with random distribution of the $B$-site cations. Despite similar ionic characters of constituent transition metals in each compound, the maximum $B$-site order attained was surprisingly high, ~90% for LCFO and ~80% for LVMO, suggesting a significant role of epitaxial stabilization in the spontaneous ordering process. Magnetization and valence state characterizations revealed that the magnetic ground state of both compounds was coincidently ferrimagnetic with saturation magnetization of ~$2\mu_B$ per formula unit, unlike those predicted theoretically. In addition, they were found to be insulating with optical band gaps of 1.6 eV and 0.9 eV for LCFO and LVMO, respectively. Our results present a wide opportunity to explore novel magnetic properties of binary transition-metal perovskites upon epitaxial stabilization of the ordered phase.

# I. INTRODUCTION

Double-perovskite oxides are an intriguing class of materials, exhibiting a number of exotic properties such as the high-Curie temperature ($T_C$) ferrimagnetism and the half-metallicity.[1] Such materials and properties can be potential candidates for a new generation of spin-based devices, thereby have been attracting renewed attention. They are expressed as $A_2B'B''O_6$, where $A$ is an alkaline- or rare-earth element and $B'$ and $B''$ are different transition-metal elements. Figure 1(a) shows a schematic structure of the double-perovskite, where the transition metals occupy the $B$-site alternately along the [111] direction to form a rock-salt-type sublattice. When 3$d$ and 4$d$ (or 5$d$) transition metals are combined, a large difference in the formal valences (FV) and ionic radii ($r_B$) permits spontaneous ordering of transition-metal ions, thus facile to synthesize in a bulk form. Representative examples are $Sr_2Fe^{3+}Mo^{5+}O_6$,[2] $Sr_2Fe^{3+}Re^{5+}O$,[3,4] and $Sr_2Cr^{3+}Re^{5+}O_6$.[5] While a few 3$d$-3$d$ combinations are known to form the ordered phase, such as $La_2Mn^{4+}B''O_6$ ($B''$ = $Fe^{2+}$, $Ni^{2+}$, $Co^{2+}$).[6,7] In any cases, the differences in FV and $r_B$ have to be large to realize the ordered phase. A boundary condition between disordered and ordered phases is suggested by Anderson $et\ al.$[8] They surveyed more than 300 compounds and summarized the crystalline phase in a FV–$r_B$ diagram, which is shown in Fig. 1(b). This phase diagram has been a useful guidance to date for studying bulk crystals or powder materials of the mixed perovskites. In fact, aforementioned $Sr_2B'B''O_6$ ($B'B''$ = $Fe^{3+}Mo^{5+}$, $Fe^{3+}Re^{5+}$, $Cr^{3+}Re^{5+}$) are located out of the disorder region (triangles), and $La_2Mn^{4+}B''O_6$ ($B''$ = $Fe^{2+}$, $Ni^{2+}$, $Co^{2+}$) will be plotted in the order region (just at right-hand side of a filled square). To the best of our knowledge, there has been no report of double-perovskite oxides in ordered forms that would be placed in the disorder region (bottom left). Here we have attempted epitaxial synthesis of $La_2Cr^{3+}Fe^{3+}O_6$ (LCFO) and $La_2V^{3+}Mn^{3+}O_6$ (LVMO) double perovskites that are marked by filled circles in Fig. 1(b). Despite vanishingly small differences in FV and $r_B$, we have succeeded in synthesize well-ordered phase of these compounds for the first time.

Our study presented here is in part motivated by seminal works by Pickett. Based on the local-spin density calculation, he showed that in LCFO the ferrimagnetic ground state with a net spin moment of $2\mu_B$ per formula unit (f.u.) is more stable than the ferromagnetic one with ~$7\mu_B$/f.u.[9] This prediction is inconsistent with the Kanamori-Goodenough (KG) rule, in which the $d^3$-$d^5$ superexchange ($Cr^{3+}/Fe^{3+}$; 3$d^3$/3$d^5$) is ferromagnetic.[10,11] On the other hand, Ueda $et\ al.$ postulated that their (111) oriented films, grown by pulsed-laser deposition (PLD) technique in a fashion of the $LaCrO_3$/$LaFeO_3$ superlattice, exhibited ferromagnetism though the measured saturation magnetization is much less than the expected value.[12,13] Until now, there is no critical evidence to support either scenario.[14,15]

Pickett has also investigated LVMO to find it as a possible half-metallic antiferromagnet (HMAFM), assuming a low-spin (LS) state of $Mn^{3+}$ ions.[9] The HMAFM is a unique class of materials having a large degree of spin polarization of conduction electrons, but vanishing macroscopic magnetic moment.[16] In the case of LVMO, local-spin moment will be exactly canceled if spins at $V^{3+}$ and $Mn^{3+}$(LS) are antiparallel ($3d^2_\downarrow 3d^4_\uparrow$; $S = -1+1 = 0$). Some experimental studies have already been performed on the bulk disordered phase of this material, but a clear conclusion is still missing because of *B*-site disorder in the material.[17]

The epitaxial synthesis of the desired materials is a trial-and-error protocol, in which one has to explore a wide region of the kinetic and/or thermodynamic landscape even if a suitable material to substrate is provided. The PLD synthesis of the double-perovskite oxides may be particularly puzzling because totally different oxidation dynamics of the transition metals must be tuned simultaneously. Such situation can be found in a growth phase diagram on $La_2VCuO_6$ (see Fig. 2).[18] The charge neutrality will be preserved by accessing two mixed valence states with either $V^{5+}Cu^{1+}$ or $V^{4+}Cu^{2+}$, while ordered phase is realized near a crossover between two oxidation states. Similar phase diagram is established for $Sr_2FeMoO_6$.[19] In this respect, tuning of oxidation dynamics is crucially important for both LCFO and LVMO systems as well.

This paper summarizes our earlier results on the epitaxial synthesis of LCFO and LVMO double-perovskite films, without using the artificial superlattice technique, and their structural and magnetic properties.[20,21] LCFO exhibits the degree of order as high as ~90% and a saturation magnetization of ~$2\mu_B$/f.u., being consistent with the Pickett's model, but violating the KG rule. LVMO is a ferrimagnet with $Mn^{3+}$ ions at a high-spin (HS) state, in contrast to the theoretically calculated one. In addition, optical properties of these double perovskites are also investigated and compared with those of a sesquioxide analog to LCFO [*i.e.*, $\alpha$-$(Cr_xFe_{1-x})_2O_3$] in order to draw new aspects of their electronic structures.

## II. EXPERIMENT
### A. Pulsed-laser deposition

A large number of LCFO and LVMO films were grown on an atomically flat surface of both-side polished (111) $SrTiO_3$ (STO) substrates by using an ultra-high vacuum PLD system.[22] KrF excimer laser pulses (5 Hz, ~1 J/cm$^2$) were focused on targets (disordered ceramic tablet, 99.99% purity) with laser spot area of 0.35 × 0.10 cm$^2$. The entire growth was *in-situ* monitored by reflection high-energy electron diffraction (RHEED). The film thickness was regulated by estimating the growth rate from periods of RHEED intensity oscillations that were observable at the initial stage when a layer-by-layer growth mode was realized. During the growth, temperature ($T_g$) and oxygen partial pressure ($P_{O2}$) were kept constants, while varied in a range of 700°C ~ 1000°C and 10 mTorr ~ 1 x 10$^{-8}$ Torr, respectively, for independent runs. After the growth, samples were cooled to room temperature, keeping $P_{O2}$ constant. As for LCFO, some of the samples were furnace-annealed in air at 400°C or 800°C for 3 h to refill residual oxygen vacancies. The film composition was analyzed by an electron probe microanalyzer (EPMA) for those grown on (001) MgO substrates under the same conditions. The composition of cations was confirmed to be identical to those of the targets regardless of growth condition. Crystalline structure was characterized by using a four-circle x-ray diffractometer (X'pert MRD, PANalytical) with CuK$\alpha$ radiation. To evaluate the degree of order accurately, x-ray diffraction (XRD) measurements were also performed at the synchrotron beamline BL-3A on the Photon Factory, KEK, Japan.

To compare with LCFO, the $\alpha$-(Cr$_x$Fe$_{1-x}$)$_2$O$_3$ (0 ≤ $x$ ≤ 1) solid-solution films were prepared.[23] The growth was performed at 740°C and $P_{O2}$ = 1 mTorr on atomically flat *c*-plane sapphire substrates in the same PLD system with a KrF excimer laser operated at 10 Hz and 0.9 J/cm$^2$. The film composition was varied with using homemade ceramics targets ($x$ = 0, 0.25, 0.50, 0.74, 1). The films having intermediate compositions were prepared by repeating alternate ablation of two neighboring targets. The composition and crystalline structure were characterized by EPMA and XRD (Rigaku, SmartLab), respectively.

**B. Characterizations**

Film thickness was regulated to be 40 ~ 80 nm for investigating the magnetic and optical properties. Magnetization measurements were performed for LCFO and LVMO films using a superconducting quantum interference device (SQUID) magnetometer (MPMS-XL, Quantum Design). The hysteresis measurements were done in a field range of ± 1 T at 5 K. The field-cooled (FC) and zero-field cooled (ZFC) magnetization were measured over a temperature range from 5 K to 300 K. For clarity, diamagnetic signal of the STO substrates was subtracted. Optical absorption spectra of LCFO and $\alpha$-(Cr$_x$Fe$_{1-x}$)$_2$O$_3$ films were evaluated from transmittance and reflectance measured with UV-Vis-IR spectroscopy at room temperature. Note that for LCFO film data at photon energy higher than 3.0 eV does not reflect actual property because of the opaque STO substrate. This is a reason why we characterize $\alpha$-(Cr$_x$Fe$_{1-x}$)$_2$O$_3$ films grown on transparent sapphire substrates to compare with the LCFO film and also reference LaCrO$_3$ and LaFeO$_3$ bulks, as will be discussed in section III-C.

## III. RESULTS AND DISCUSSION
### A. Epitaxial growth and structural characterizations

The ordered films were reproducibly obtained for both LCFO and LVMO grown in a wide $T_g$–$P_{O2}$ range. We have mapped the *B*-site order as a function of $T_g$ and $P_{O2}$ to find the best condition at $T_g$ = 1000°C (880°C) and $P_{O2}$ = 1 x 10$^{-4}$ Torr (1 x 10$^{-5}$ Torr) for LCFO (LVMO), where the Cr/Fe (V/Mn) order reached ~90% (~80%).[20,21] Figures 3(a) and 3(b) show $\theta$–2$\theta$ XRD patterns along the out-of-plane reflections for LCFO and LVMO films, respectively. Superlattice reflections can be seen at near half-integers *L* (a reciprocal lattice unit referred to STO). These results present striking demonstration of spontaneous ordering, which is unexpected in a bulk form taking similar ionic characters of constituent transition-metal ions into account.

We found that these films were pseudomorphically grown on (111) STO, reflecting small lattice mismatches to cubic STO (3.905 Å); pseudo-cubic lattice constants were evaluated to be 3.915 Å and 3.919 Å for LCFO and LVMO, respectively. Atomically abrupt and coherent interface was clearly observed for the LCFO/STO by using high-angle annular dark-field scanning transmission electron microscopy.[20] The valence state of Cr and Fe ions was confirmed to be both 3+ by electron-energy loss spectroscopy analysis. Also, the valence states of V$^{3+}$ and Mn$^{3+}$ were identified by synchrotron radiation photoemission spectroscopy. In addition, a HS state of Mn$^{3+}$ ions was deduced from the photoemission and an optical absorption spectrum.[21]

The mechanism of spontaneous ordering is unclear at the moment, yet we find a general tendency that lattice disorder corrupts *B*-site ordering, namely, a degree of ordering systematically decreased as rocking curve width for a fundamental perovskite reflection increased.[20] This epitaxy-induced effect with a non-equilibrium nature of PLD growth could lead to spontaneous atomic ordering in the present and other systems that are inaccessible in bulk materials. In this respect, our study presents a new benchmark beyond a framework of conventional notion of epitaxial stabilization.[24]

## B. Magnetic properties

The as-grown LCFO films exhibited vanishingly small magnetization as will be discussed later. As for the oxygen-annealed films (400°C), however, the in-plane magnetization loop taken at 5 K traced well-defined hysteresis [Fig. 4(a)]. The magnitude of saturation magnetization ($M_s$) was ~$2\mu_B$/f.u. being consistent with antiferromagnetic ordering of local-spin moment, namely ferrimagnetic [Cr($3d^3_\downarrow$)Fe($3d^5_\uparrow$); $S = -3/2+5/2 = 1$]. The temperature dependence of FC magnetization showed a clear magnetic transition at 45 K [Fig. 4(b)], which was deduced from the intercept of a liner fit to $1/M(T)$.[20] Note that the sample annealed at higher temperature (800°C) showed structural and magnetic properties almost identical to those of the 400°C-annealed one, suggesting that the ordered double-perovskite form is stable against thermally assisted cation redistribution.

The in-plane magnetization loop measured for the as-grown LVMO films was similar to that of annealed LCFO films, but the coercive field was smaller [Fig. 5(a)]. Note that oxygen annealing was not conducted to maintain the trivalent state of V ions. The magnitude of $M_s$ was ~$1.9\mu_B$/f.u. being consistent with antiferromagnetic ordering of local-spin moment with $V^{3+}$ and $Mn^{3+}$ ions at a HS state ($3d^2_\downarrow 3d^4_\uparrow$; $S = -1+2 = 1$). Figure 5(b) shows temperature dependent FC and ZFC magnetization curves taken under 0.1 T and 0.4 T. At a glance, a bifurcation in FC and ZFC curves is reminiscent of a spin-glass magnet. But this observation arises from coercivity. Indeed, the coercive field vanished at a Néel temperature of $T_N = 21$ K, where bifurcation originated.[21]

We notice that the observed magnetic transition temperatures for LCFO and LVMO are considerably lower than $T_N$ of the parent compounds (280 K for $LaCrO_3$, 750 K for $LaFeO_3$, 140 K for both $LaVO_3$ and $LaMnO_3$),[25] suggesting relatively weak antiferromagnetic superexchange. According to the KG rule, the $pd\sigma$ hybridization between the $e_g$ orbitals of the transition-metal ions and the oxygen $p\sigma$ orbital is ferromagnetic, while the $pd\pi$ hybridization with the $t_{2g}$ orbitals is antiferromagnetic. The amount of orbital overlap is larger for $pd\sigma$ hybridization, while the number of paths is larger for $pd\pi$ hybridization. This competition would lead to instability in any of dominate one for the $d^3$-$d^5$ superexchange.[14] We anticipate that this mechanism is also responsible for the $d^2$-$d^4$ superexchange in LVMO.

Next let us discuss the dependence of $M_s$ on degree of order in LCFO. Here we wish to employ the notion of antisite defect (AS) instead of the Cr/Fe order, which is defined as the percentage of misplaced Cr at Fe site and vice versa, thus a single-perovskite phase with complete *B*-site disorder would be denoted by *AS* = 0.5.[26] Figure 6(a) shows *AS* dependence of $M_s$ for as-grown and annealed samples. Note that oxygen annealing led to a drastic enhancement of low-temperature magnetization. We suggest that residual oxygen vacancies existed in the as-grown samples, leading to suppression of antiferromagnetic ordering of local-spin moment. The magnitude of $M_s$ systematically decreased with increasing *AS* and vanished at *AS* ~ 0.37 for both cases. This systematic dependence can be explained as follows. Considering that type of antiferromagnetic ordering in LaCrO$_3$ and LaFeO$_3$ is *G*-type, the spin moment at the antisite will be antiparallel with respect to the magnetization of the host, as depicted in Fig. 6(b). In this case, each misplaced Cr ion reduces $M_s$ by 2$\mu_B$ ($3d^5_\uparrow \rightarrow 3d^3_\uparrow$; *S* = –5/2+2/3 = –1), while each misplaced Fe ion also reduces $M_s$ by 2$\mu_B$ ($3d^3_\downarrow \rightarrow 3d^5_\downarrow$; *S* = 3/2–5/2 = –1), giving a loss of the spin magnetization per antisite of 4$\mu_B$. Then, one expects that $M_s$ varies as (2–4*AS*)$\mu_B$/f.u. and thus vanishes at *AS* = 0.5. In reality, the other factors such as magnetic frustration or long-range interaction may disturb the nearest- and the next-nearest-neighbor superexchange,[27] which is consistent with rather rapid damping of $M_s$, as seen here experimentally.

### C. Electronic structures

Figure 7(a) shows absorption spectrum of a LCFO film in a comparison with optical conductivity spectra of the parent compounds.[28] It is remarkable that absorption edge of LCFO exhibited red-shift with respect to LaCrO$_3$ and LaFeO$_3$. The optical band gap is evaluated to be approximately 1.6 eV through the linear interpolation. In order to reveal the origin of the red-shift, we have measured the absorption spectra for $\alpha$-(Cr$_x$Fe$_{1-x}$)$_2$O$_3$ films grown on (0001) $\alpha$-Al$_2$O$_3$,[23] which are shown in Fig. 7(b). We have found that the spectral weight around 2.0 eV dramatically increased as the composition approached to *x* = 0.5. This lower-energy absorption band can be attributed to excitation of valence electrons at Cr $t_{2g}$ and/or O $2p$ orbitals to the empty Fe $t_{2g}$ orbital.[29] Given a common coordination in the octahedral crystal field, the same mechanism is likely responsible for the observed red-shift in LCFO film.

We have also measured absorption spectrum of a LVMO film to find that its absorption edge (0.9 eV) was close to that of LaMnO$_3$ (1.0 eV).[21] In addition, synchrotron radiation photoemission spectroscopy measurements for LVMO revealed that the valence band maximum is consisted of a narrow state arising from partially occupied V $t_{2g}$ and Mn $e_g$ orbitals and a broad O $2p$ state. Based on these results, we concluded that the lowest transition occurs as valence electrons at Mn $e_g$ and/or O $2p$ orbitals are excited to the empty Mn $e_g$ orbital (note that an interatomic transition from Mn $e_g$ to the empty V orbitals is spin-forbidden.).

Finally, we wish to present electronic structures of LCFO and LVMO, being consistent with all of our experimental results [Figs. 8(a) and 8(b), respectively]. It is worth mentioning that the electronic structures of LCFO and LVMO are slightly modified from, but preserving characteristics of parent compounds, LaCrO$_3$ (mixed type of insulator), LaFeO$_3$, LaMnO$_3$ (charge transfer insulators), and LaVO$_3$ (Mott-Hubbard insulator).[28] To summarize the magnetic and electronic properties, our experimental results are compared to the results of local-spin density calculation in Table I.

## IV. CONCLUSIONS

Two novel double-perovskite oxides have been synthesized by epitaxial stabilization through pulsed-laser deposition on (111) STO. As for LCFO, saturation magnetization reached ~2$\mu_B$/f.u. at 5 K as the Cr/Fe order increases up to 90%, from which the ground-state magnetic order has been verified to be ferrimagnetic ($3d^3 3d^5$; $S = -3/2 + 5/2 = 1$) below 45 K. LVMO (V/Mn order ~80%, $T_N = 21$ K) is also ferrimagnetic with Mn$^{3+}$ at a HS state ($3d^2 3d^4$; $S = -1 + 2 = 1$). Both compounds are insulators exhibiting charge transfer gaps in visible light region. $\alpha$-(Cr$_x$Fe$_{1-x}$)$_2$O$_3$ as a sesquioxide analog to LCFO also shows similar optical properties including smaller band gaps than its end members. Based on these results, we propose plausible band structures of La$_2$CrFeO$_6$ and La$_2$VMnO$_6$. Although it remains an open question why epitaxial films yield double-perovskite phases while bulk does not, our study sheds light on physical properties of the double perovskites, consisting of earth-abundant transition metals and yet experimentally unexplored.

## ACKNOWLEDGMENTS

The Authors thank Dr. D. Okuyama, Prof. R. Kumai, Prof. T. Arima, and Prof. Y. Tokura for synchrotron radiation XRD measurements and structural analysis, Dr. M. Saito, Prof. S. Tsukimoto, and Prof. Y. Ikuhara for STEM and EELS analyses, Prof. H. Kumigashira, Dr. K. Yoshimatsu, Dr. E. Sakai, Prof. M. Oshima, Dr. T. Makino, and Dr. Y. Kozuka for synchrotron radiation XPS and optical spectroscopy analyses, and all of them for fruitful discussions. A.O., H.M., and T.O. thank Prof. Y. Wada and Dr. D. Mochizuki for assistance regarding EPMA analysis. This work was partly supported by a Grant-in-Aid for Scientific Research (No.

24245035) from the Japan Society for the Promotion of Science. H. M. is supported by the Program for Leading Graduate Schools "Academy for Co-creative Education of Environment and Energy Science", of the Ministry of Education, Culture, Sports, Science and Technology.

**REFERENCES**

1. W. E. Pickett and J. S. Moodera: Half metallic magnets. *Physics Today* **54**(5), 39 (2001).
2. K.-I. Kobayashi, T. Kimura, H. Sawada, K. Terakura, and Y. Tokura: Room-temperature magnetoresistance in an oxide material with an ordered double-perovskite structure. *Nature* **395**, 677 (1998).
3. K.-I. Kobayashi, T. Kimura, Y. Tomioka, H. Sawada, K. Terakura, and Y. Tokura: Intergrain tunneling magnetoresistance in polycrystals of the ordered double perovskite $Sr_2FeReO_6$. *Phys. Rev. B* **59**, 11159 (1999).
4. T. H. Kim, M. Uehara, S.-W. Cheong, and S. Lee: Large room-temperature intergrain magnetoresistance in double perovskite $SrFe_{1-x}(Mo$ or $Re)_xO_3$. *Appl. Phys. Lett.* **74**, 1737 (1999).
5. H. Kato, T. Okuda, Y. Okimoto, Y. Tomioka, Y. Takenoya, A. Ohkubo, M. Kawasaki, and Y. Tokura: Metallic ordered double-perovskite $Sr_2CrReO_6$ with maximal Curie temperature of 635 K. *Appl. Phys. Lett.* **81**, 328 (2002).
6. G. Blasse: Ferromagnetic interaction in non-metallic perovskites. *J. Phys. Chem. Solids* **26**, 1969 (1965).
7. K. Ueda, Y. Muraoka, H. Tabata, and T. Kawai: Atomic ordering in the $LaFe_{0.5}Mn_{0.5}O_3$ solid solution film. *Appl. Phys. Lett.* **78**, 512 (2001).
8. M. T. Anderson, K. B. Greenwood, G. A. Taylor, and K. R. Poeppelmeier: B-cation arrangements in double perovskites. *Prog. Solid State Chem.* **22**, 197 (1993).
9. W. E. Pickett: Spin-density-functional-based search for half-metallic antiferromagnets. *Phys. Rev. B* **57**, 10613 (1998).
10. J. Kanamori: Superexchange interaction and symmetry properties of electron orbitals. *J. Phys. Chem. Solids* **10**, 87 (1959).
11. J. B. Goodenough: Theory of the role of covalence in the perovskite-type manganites [La, *M*(II)]$MnO_3$. *Phys. Rev.* **100**, 564 (1955).
12. K. Ueda, H. Tabata, and T. Kawai: Ferromagnetism in $LaFeO_3$-$LaCrO_3$ superlattices. *Science* **280**, 1064 (1998).

13. W. E. Pickett, G. I. Meijer, K. Ueda, H. Tabata, and T. Kawai: Ferromagnetic superlattices. *Science* **281**, 1571a (1998).

14. K. Miura and K. Terakura: Electronic and magnetic properties of La$_2$FeCrO$_6$: Superexchange interaction for a $d^5$-$d^3$ system. *Phys. Rev. B* **63**, 104402 (2001).

15. B. Gray, H. N. Lee, J. Liu, J. Chakhalian, and J. W. Freeland: Local electronic and magnetic studies of an artificial La$_2$FeCrO$_6$ double perovskite. *Appl. Phys. Lett.* **97**, 013105 (2010).

16. H. van Leuken and R. A. de Groot: Half-metallic antiferromagnets. *Phys. Rev. Lett.* **74**, 1171 (1995).

17. J. Androulakis, N. Katsarakis, and J. Giapintzakis: Realization of La$_2$MnVO$_6$: search for half-metallic antiferromagnetism? *Solid State Commun.* **124**, 77 (2002).

18. S. Chakraverty, A. Ohtomo, and M. Kawasaki (unpublished).

19. T. Manako, M. Izumi, Y. Konishi, K.-I. Kobayashi, M. Kawasaki, and Y. Tokura: Epitaxial thin films of ordered double perovskite Sr$_2$FeMoO$_6$. *Appl. Phys. Lett.* **74**, 2215 (1999).

20. S. Chakraverty, A. Ohtomo, D. Okuyama, M. Saito, M. Okude, R. Kumai, T. Arima, Y. Tokura, S. Tsukimoto, Y. Ikuhara, and M. Kawasaki: Ferrimagnetism and spontaneous ordering of transition metals in double perovskite La$_2$CrFeO$_6$ films. *Phys. Rev. B* **84**, 064436 (2011).

21. S. Chakraverty, K. Yoshimatsu, Y. Kozuka, H. Kumigashira, M. Oshima, T. Makino, A. Ohtomo, and M. Kawasaki: Magnetic and electronic properties of ordered double-perovskite La$_2$VMnO$_6$ thin films. *Phys. Rev. B* **84**, 132411 (2011).

22. S. Chakraverty, A. Ohtomo, M. Okude, K. Ueno, and M. Kawasaki: Epitaxial structure of (001)- and (111)-oriented perovskite ferrate films grown by pulsed-laser deposition. *Cryst. Growth Des.* **10**(4), 1725 (2010).

23. H. Mashiko, T. Oshima, and A. Ohtomo: Band-gap narrowing in $\alpha$-(Cr$_x$Fe$_{1-x}$)$_2$O$_3$ solid-solution films. *Appl. Phys. Lett.* **99**, 241904 (2011).

24. O. Yu. Gorbenko, S. V. Samoilenkov, I. E. Graboy, and A. R. Kaul: Epitaxial stabilization of oxides in thin films. *Chem. Mater.* **14**(10), 4026 (2002).

25. M. Imada, A. Fujimori, and Y. Tokura: Metal-insulator transitions. *Rev. Mod. Phys.* **70**, 1039 (1998).

26. D. Serrate, J. M. De Teresa, and M. R. Ibarra: Double perovskites with ferromagnetism above room temperature. *J. Phys.: Condens. Matter* **19**, 023201 (2007).

27. D. Serrate, J. M. De Teresa, and M. R. Ibarra: Double perovskites with ferromagnetism above room temperature. *J. Phys.: Condens. Matter* **19**, 023201 (2007).
28. T. Arima, Y. Tokura, and J. B. Torrance: Variation of optical gaps in perovskite-type 3$d$ transition-metal oxides. *Phys. Rev. B* **48**, 17006 (1993).
29. H. S. Nabi and R. Pentcheva: Energetic stability and magnetic coupling in $(Cr_{1-x}Fe_x)_2O_3$: Evidence for a ferrimagnetic ilmenite-type superlattice from first principles. *Phys. Rev. B* **83**, 214424 (2011).

**TABLE I.** A comparison between experimental and theoretical calculation results on the magnetic properties of $La_2CrFeO_6$ and $La_2VMnO_6$ (FM: ferromagnetic, FiM: ferrimagnetic, HFAFM: half-metallic antiferromagnetic).

| System | 3d-electron-configuration | | Order | $M_s$ ($\mu_B$/f.u.) | Ref. |
|---|---|---|---|---|---|
| $La_2CrFeO_6$ | | | | | |
| Theory | $Cr^{3+}$ $(t_{2g})^3$ | $Fe^{3+}$ $(t_{2g})^3(e_g)^2$ | FM | 7.15 | 9,14 |
| Theory | $Cr^{3+}$ $(t_{2g})^3$ | $Fe^{3+}$ $(t_{2g})^3(e_g)^2$ | FiM | 2 | 9 |
| Exp. | $Cr^{3+}$ $(t_{2g})^3$ | $Fe^{3+}$ $(t_{2g})^3(e_g)^2$ | FiM | 2 | 20 |
| $La_2VMnO_6$ | | | | | |
| Theory | $V^{3+}$ $(t_{2g})^2$ | $Mn^{3+}$ $(t_{2g})^4(e_g)^0$ | HFAFM | 0 | 9 |
| Exp. | $V^{3+}$ $(t_{2g})^2$ | $Mn^{3+}$ $(t_{2g})^3(e_g)^1$ | FiM | 2 | 21 |

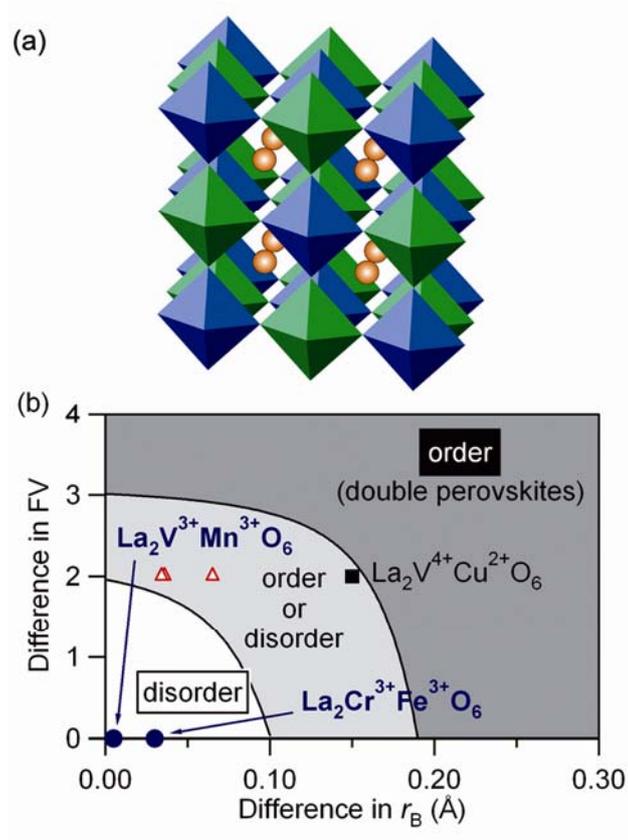

**FIG. 1.** (Color online) (a) Schematic unit lattice of double-perovskite $A_2B'B''O_6$. The transition metals $B'$ and $B''$ surrounded with six-coordinated oxygen ions (represented by blue and green octahedral, respectively) form a rock-salt-type sublattice. (b) Bulk phase diagram as a function of differences in formal valences (FV) and ionic radii ($r_B$) (adopted from Ref. 8). The double-perovskite oxides investigated in this study are marked by filled circles and square. Open triangles represent high-$T_C$ half-metallic ferrimagnets, $Sr_2B'B''O_6$ ($B'B''$ = CrRe, FeMo, FeRe, from left to right in the figure).

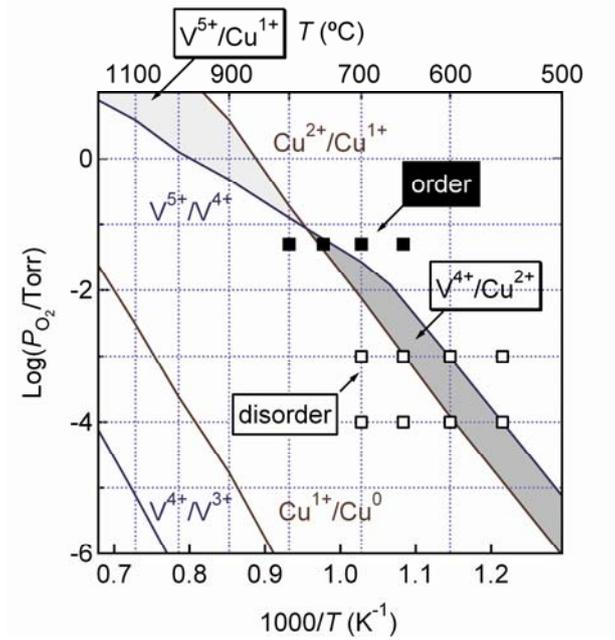

**FIG. 2.** (Color online) Epitaxial growth phase diagram as a function of $P_{O2}$ and growth temperature for $La_2VCuO_6$ on (111) STO. The ordered and disordered phases are marked by filled and open squares, respectively. Solid lines and shaded regions indicate phase equilibrium of V–O and Cu–O binary systems and a stable condition for coexistence of $V^{4+}/Cu^{2+}$ (dark) or $V^{5+}/Cu^{1+}$ (light), respectively.

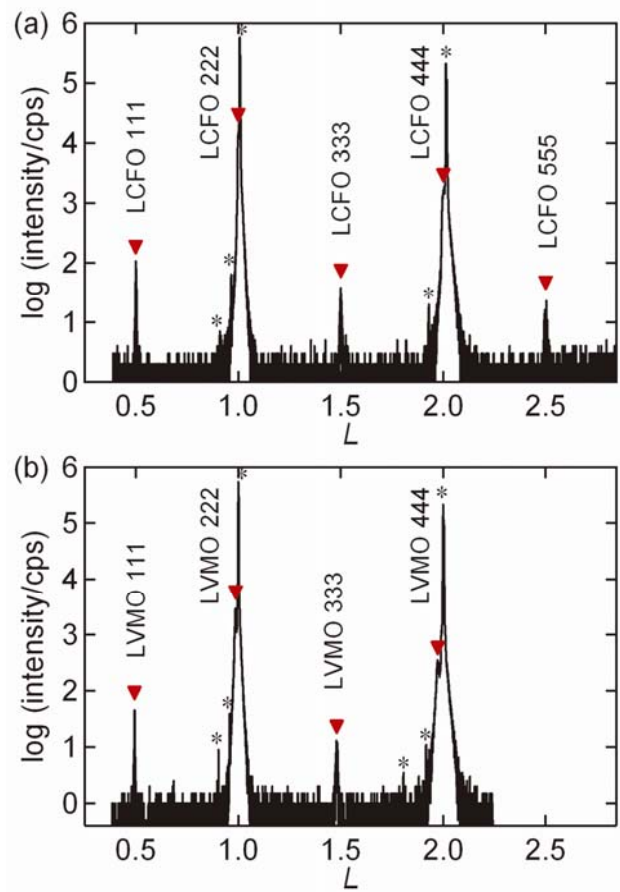

**FIG. 3.** (Color online) X-ray diffraction patterns of the highest ordered (a) LCFO and (b) LVMO films grown on (111) STO. Film and substrate peaks are marked with triangles and asterisks, respectively. The superlattice reflections correspond to odd-index peaks located at near half-integers *L*.

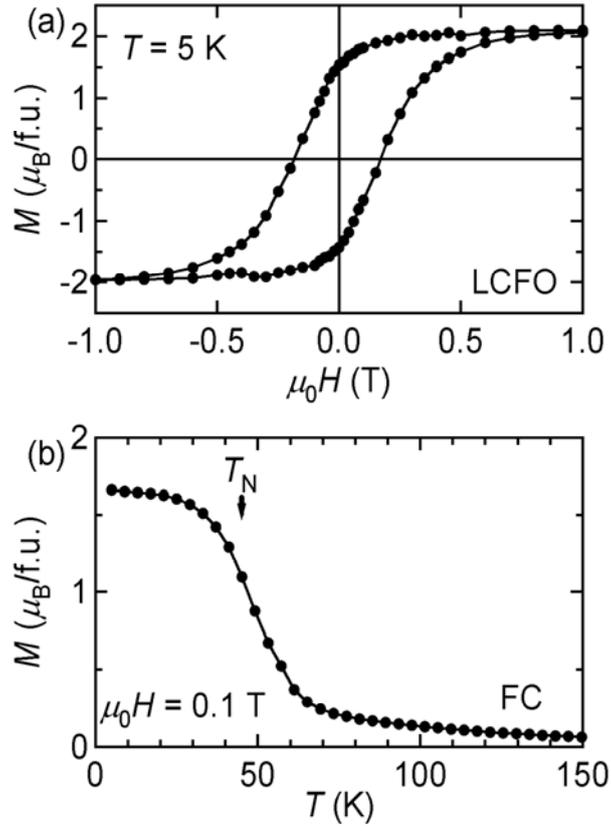

**FIG. 4.** (a) Magnetization hysteresis curve taken at 5 K for oxygen-annealed LCFO film (88% ordered). (b) The temperature dependence of FC magnetization taken during warming under 0.1 T for the same sample.

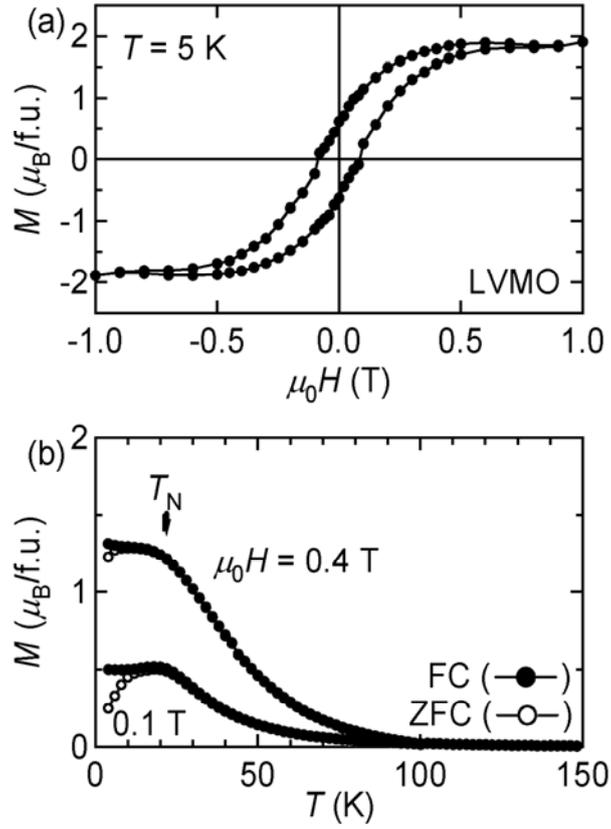

**FIG. 5.** (a) Magnetization hysteresis curve taken at 5 K for LVMO film (80% ordered). (b) The temperature dependence of FC and ZFC magnetization taken during warming under 0.1 T and 0.4 T for the same sample.

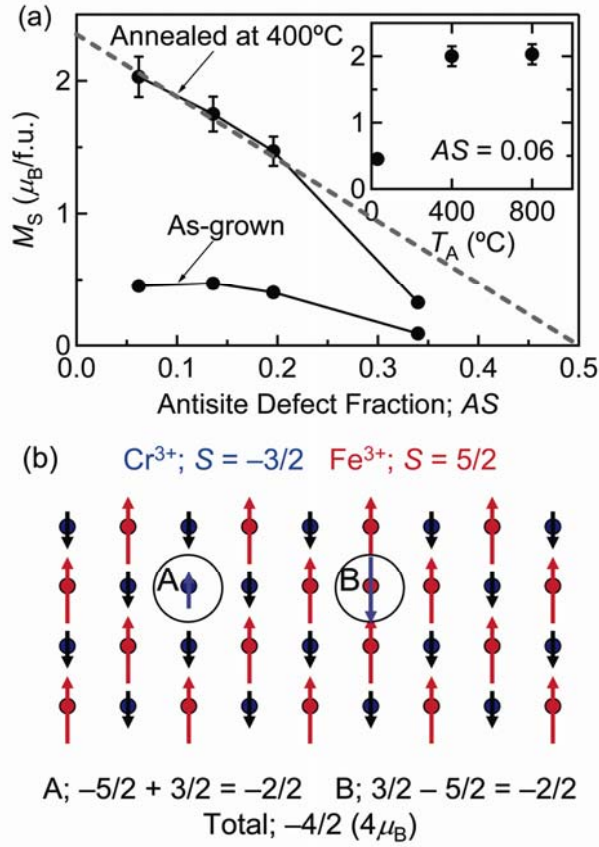

**FIG. 6.** (Color online) (a) Saturation magnetization ($M_s$) as a function of antisite defect fraction (*AS*) for as-grown and oxygen-annealed LCFO films. Inset depicts $M_s$ as a function of annealing temperature ($T_A$) for *AS* = 0.06 sample. The error bars reflect inaccuracy of the sample volume. Broken line indicates dependence expected from a model in (b). (b) Model of arrangement of local-spin moment at an antisite defect [a pair of $Cr_{Fe}$ (circle A) and $Fe_{Cr}$ (circle B)], showing net spin compensation of $4\mu_B$ per site. Note that the spin direction at each antisite defect is opposite to either that at the right positions or that at the nearest-neighbor sites and overall antiferromagnetic ordering of the local-spin moment is preserved.

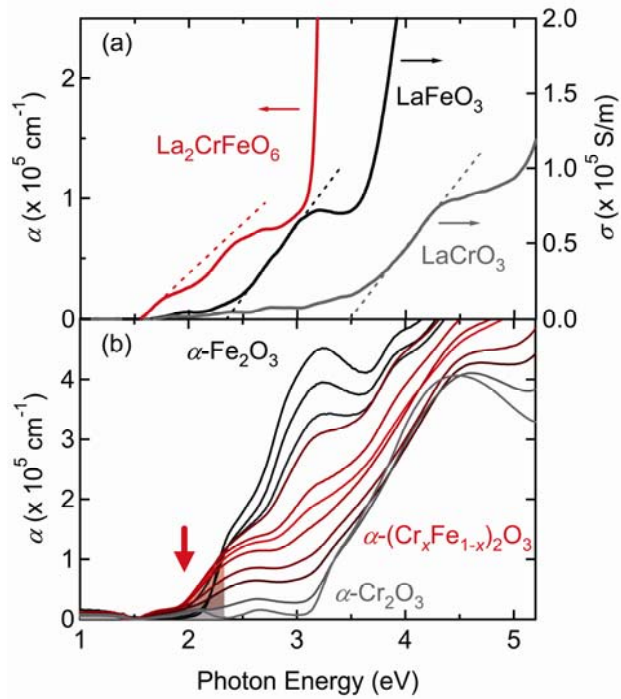

**FIG. 7.** (Color online) (a) Optical absorption spectrum of LCFO film taken at room temperature and optical conductivity spectra of parent compounds, $LaCrO_3$ and $LaFeO_3$ bulks.[28] (b) Optical absorption spectra of $\alpha$-$(Cr_xFe_{1-x})_2O_3$ films ($x$ = 0, 0.05, 0.15, 0.25, 0.4, 0.5, 0.6, 0.74, 0.85, 0.95, and 1).[23] Arrow and shaded region indicate absorption band assignable to a transition from O $2p$/Cr $3d(t_{2g})$ to Fe $3d(t_{2g})$.

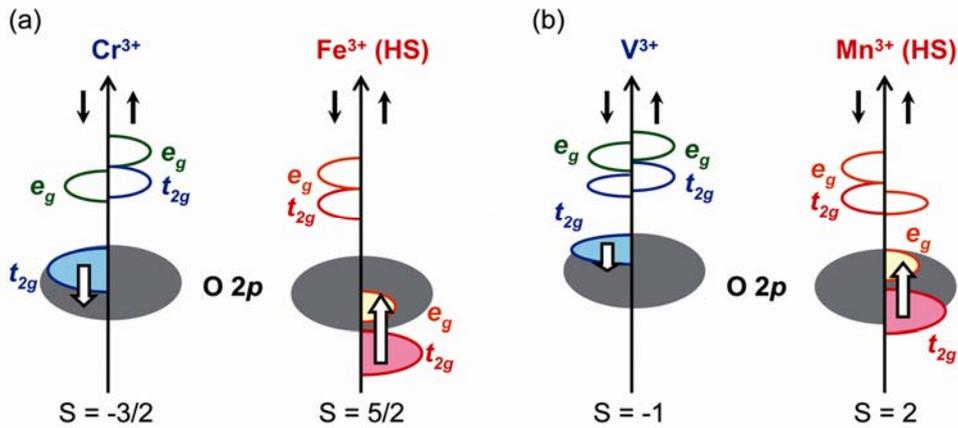

**FIG. 8.** (Color online) Schematic band structures of (a) a $Cr^{3+}$–$Fe^{3+}$–O compound and (b) a $V^{3+}$–$Mn^{3+}$(HS) –O compound having antiferromagnetic spin ordering. Energy levels of empty bands are tentatively assigned.

19